\documentclass[a4paper,11pt,dvitex]{article}
\usepackage{pos}

\title{Particle density probability distribution function and center symmetry breaking in finite density lattice gauge theories}
\ShortTitle{Particle density probability distribution function and center symmetry breaking}

\author*[a]{Shinji Ejiri}
\affiliation[a]{Department of Physics, Niigata University, 
Niigata 950-2181, Japan}

\emailAdd{ejiri@muse.sc.niigata-u.ac.jp}
\abstract{
We study the nature of the phase transition at high temperature and high density in lattice gauge theories by focusing on the probability distribution function, which represents the probability that a certain density will be realized in a heat bath. 
The probability distribution function is obtained by constructing a canonical partition function by fixing the number of particles from the grand partition function. 
However, if the $Z_3$ center symmetry, which is important for understanding the finite temperature phase transition of $SU(3)$ lattice gauge theory, is maintained on a finite lattice, the probability distribution function is always zero, except when the number of particles is a multiple of 3. 
For $U(1)$ gauge theory, this problem is more extreme. 
The probability distribution becomes zero when the particle number is not zero.
In this study, we find a solution to this problem and propose a method of avoiding the sign problem, which is an important problem at finite density, using the center symmetry. 
This problem is essentially the same as the problem that the expectation value of the Polyakov loop is always zero when calculating with finite volume. 
In the case of $U(1)$ lattice gauge theory with heavy fermions, numerical simulations are actually performed, and we demonstrate that the probability distribution function at a finite density can be calculated by the method proposed in this study.
}

\FullConference{%
 The 38th International Symposium on Lattice Field Theory, LATTICE2021
  26th-30th July, 2021
  Zoom/Gather@Massachusetts Institute of Technology
}

\begin{document}
\maketitle

\section{Introduction}
\label{intro}

Theoretical study of the probability distribution of particle density is important for investigating the aspect of density fluctuation in the heat bath generated by heavy ion collision experiments and finding the critical point at finite density.
We develop a method for calculating the probability distribution function by the first-principle calculation of lattice QCD.
The probability distribution function is obtained by constructing a canonical partition function by fixing the number of particles from the grand partition function. 
The relation between the grand partition function $Z_{GC}(T, \mu)$ with the chemical potential $\mu$ and the canonical partition function $Z_C(T,N)$ with the particle number $N$ is given by the fugacity expansion,
\begin{eqnarray}
Z_{GC}(T,\mu) = \sum_{N} Z_C(T,N) e^{N \mu/T}.
\label{eq:fex}
\end{eqnarray}
The left-hand side of this equation $Z_{GC}(T, \mu)$ is the normalization factor of the Boltzmann weight, and is classified by $N$ in the right-hand side.
Hence, $Z_C(T,N) e^{N \mu/T}$ can be regarded as a weight factor for each $N$, and the probability distribution $W(N)$ is in proportion to $Z_C(T,N) e^{N \mu/T}$.
Thus, the effective potential of $N$ can be defined as $- \ln W(N) = - \ln Z_C (T,N) - N \mu/T$.

However, if the $Z_3$ center symmetry, which spontaneously breaks in the deconfinement phase transition of $SU(3)$ lattice gauge theory, is strictly maintained on a finite lattice, the canonical partition function of QCD $Z_C (T,N)$ is zero when the number of particles is not a multiple of 3. 
Moreover, in the case of $U(1)$ lattice gauge theory, the situation is extreme.
The partition function $Z_C (T,N)$ is exactly zero for $N \neq 0$.
This means that the existence probability of charged particles is zero, which is unacceptable.
To solve this problem, we focus on $U(1)$ lattice gauge theory.
In the calculation of $Z_C$, it is essential to break the center symmetry adding a small external field.
Then, we find that the central symmetry can be used to avoid sign problems.

In the next section, we explain the canonical partition function when the theory has the center symmetry.
We calculate the canonical partition function for $U(1)$ gauge theory with heavy dynamical fermions in Sec.~\ref{canonical}.
We also propose a method to avoid the sign problem using the $U(1)$ center symmetry.
Our conclusions are given in Sec.~\ref{summary}

\section{Center symmetry breaking in $U(1)$ gauge theory}
\label{breaking}

\begin{figure}[tb]
\centering
\vspace{-3mm}
\includegraphics[width=6.5cm,clip]{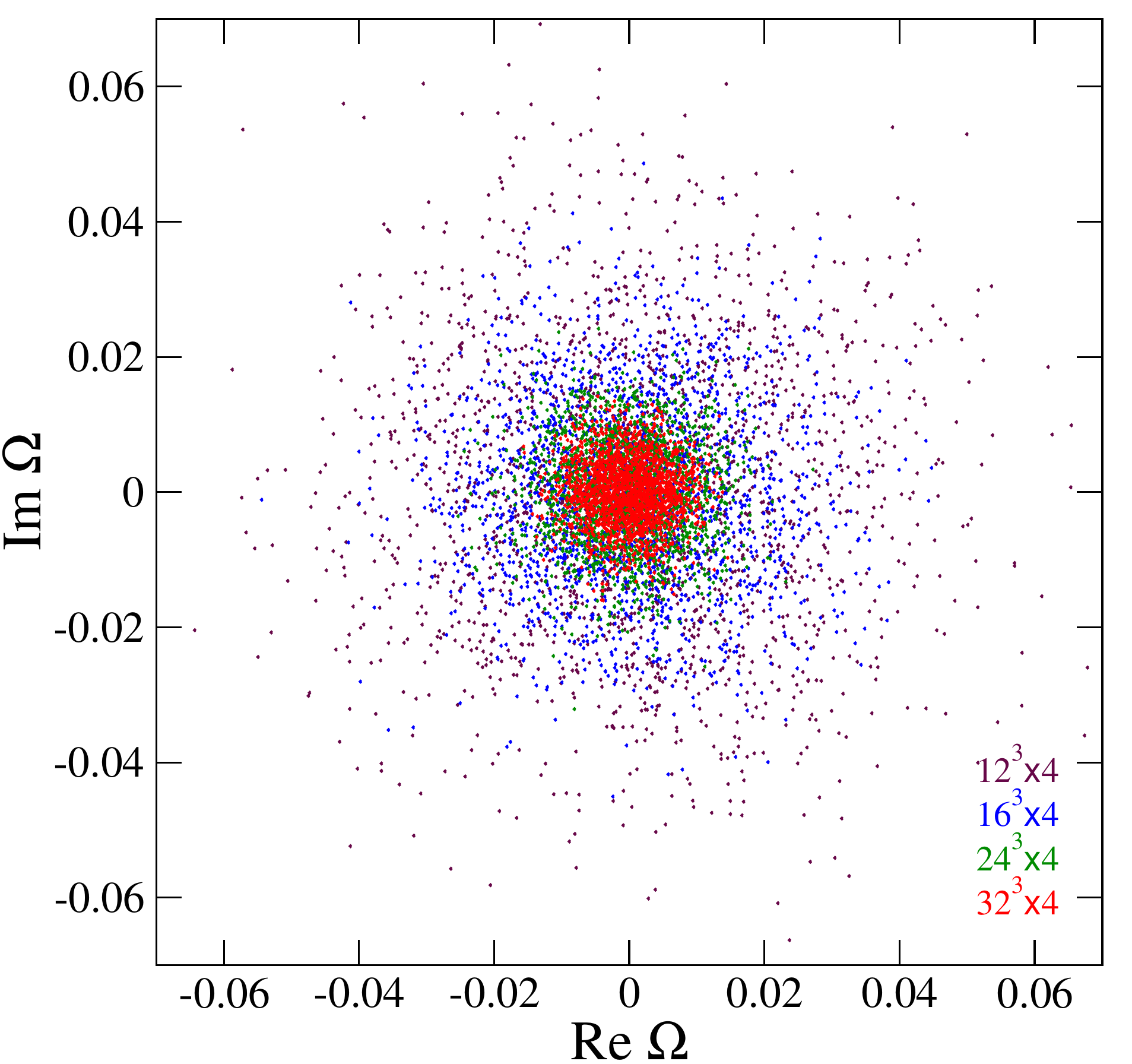}
\hspace{5mm}
\includegraphics[width=6.5cm,clip]{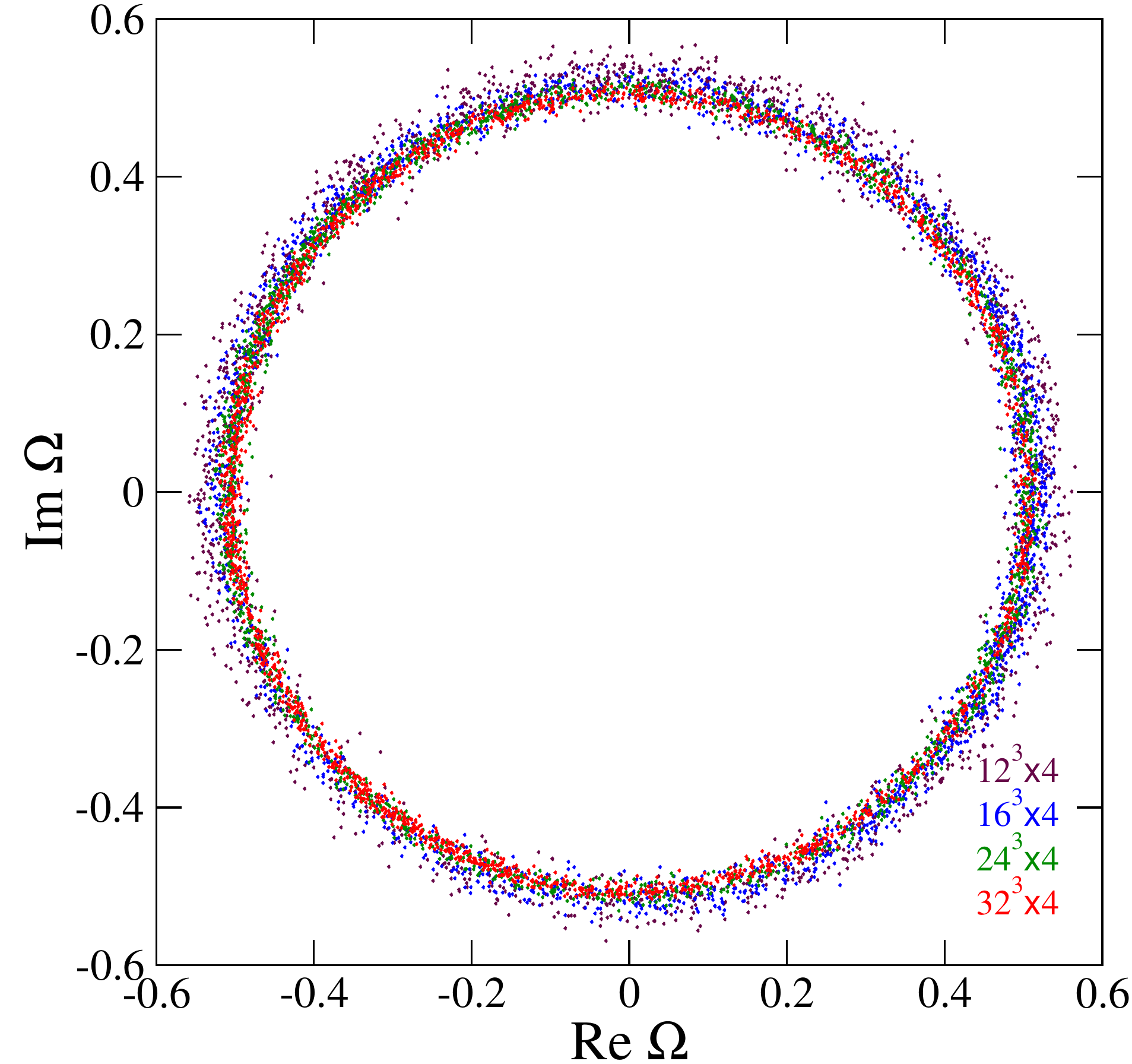}
%\vspace{-5mm}
\caption{Distribution of Polyakov loops in the complex plane generated by simulations of $U (1)$ lattice gauge theory at $\beta=0.90$ (left) and 1.10 (right). The temporal lattice size $N_t$ is 4.
Purple, blue, green and red symbols are the results of the spatial lattice size $N_s=12$, 16, 24 and 32, respectively.
}
\label{fig1}
\end{figure}

The grand partition function is 
\begin{eqnarray}
Z_{GC}(T,\mu) = \int {\cal D} U (\det M(\kappa, \mu)) ^{N_{\rm f}} \ e^{-S_g}
\end{eqnarray}
for the $N_{\rm f}$ degenerate flavor case. Here, $S_g$ is the gauge action and $M(\kappa, \mu)$ is the fermion kernel.
The fermion determinant $\det M$ can be expressed as the sum of Wilson loops.
We perform a hopping parameter expansion of $\ln \det M$. 
Then, the $n^{\rm th}$-order Taylor expansion coefficients are given by  the sum of $n$-step connected Wilson loops 
\cite{Saito:2011fs,Saito:2013vja,Ejiri:2019csa,Kiyohara:2021smr,Wakabayashi:2021eye}. 
When the hopping parameter $\kappa$ is small, the leading order contribution consists of the plaquette $P$, and the Polyakov loop $\Omega$ \cite{Saito:2013vja,Wakabayashi:2021eye}: 
\begin{eqnarray}
\ln \det M (\kappa, \mu) 
= 96 N_{\rm site} \kappa^4 P 
+ 2^{N_t+1}N_s^3 \left( \kappa^{N_t} e^{\mu/T} \Omega
+ \kappa^{N_t} e^{\mu/T} \Omega^* 
\right) + \cdots 
\label{eq:proddetM}
\end{eqnarray}
for the standard Wilson fermion on a lattice with the spatial size $N_s$ and temporal size $N_t$.
At finite density with the chemical potential $\mu$, the $\mu$-dependence of each Wilson loop term, which is wound $m$ times by the anti-periodic boundary condition in the time direction, appears as a factor $e^{m \mu/T}$ 
because the hopping term in the time direction is $(1-\gamma_4) U_{x,4} e^{\mu a}$ or
$(1+\gamma_4) U_{x,4}^{\dagger} e^{-\mu a}$ and $(e^{\mu a})^{N_t} =e^{\mu/T}$, where $U_{x,4}$ is a link variable and $a$ is the lattice spacing.
We classify these expansion terms by the winding number $m$.
If we reconstruct the expansion of $\ln \det M$ into the expansion of $\det M$, $Z_{GC}$ is expressed in the form of the fugacity expansion, Eq.~(\ref{eq:fex}).
Therefore, the fugacity expansion is a winding number expansion.

Under the $U(1)$ center transformation: the time components of all link variables in one time slice are changed as
$U_{(\vec{x},t),4} \to e^{i \theta} U_{(\vec{x},t),4}$, \ 
$Z_C(T,N)$ changes as 
\begin{eqnarray}
Z_C (T,N) \to e^{iN \theta} Z_C(N,T), 
\end{eqnarray}
since $Z_C$ is multiplied by $e^{i \theta}$ $N$ times at the time slice in total.
Because $S_g$ and the integral measure are invariant, $Z_C (T,N) = e^{iN \theta} Z_C(N,T)$. 
Thus, the canonical partition function is zero, 
\begin{eqnarray}
Z_C (T,N) = \frac{1}{2 \pi} \int_0^{2 \pi} e^{iN \theta} Z_C(N,T) d \theta
= 0, 
\end{eqnarray}
except for $N=0$.
This means that there cannot exist particles that interact with the gauge field.
In the confined phase, the probability of existence of charged particles is zero because of the confinement.
However, charged particles must be exist in the deconfinement phase. 
$Z_C (T,N)=0$ is unacceptable.
In such a case, the grand partition function $Z_{GC}(T,\mu)$ is equal to $Z_C(T,0)$, 
and $Z_{GC}$ does not depend on the chemical potential.

This problem is a common problem when calculating the expectation value of the order parameter for investigating spontaneous symmetry breaking.
As seen in Eq.~(\ref{eq:proddetM}), the first nontrivial term of $\ln \det M$ is proportional to $\Omega \ e^{\mu/T}$ when $\kappa$ is small.
Thus, the canonical partition of $N=1$ with a small $\kappa$ is proportional to $\langle \Omega \rangle$, where the expectation value is computed in a quenched simulation.
We perform quenched QCD simulations.
Figure~\ref{fig1} shows the distribution of the Polyakov loop in the complex plane.
The left panel shows the result of the confinement phase (symmetric phase) at an inverse gauge coupling $\beta =1/g^2 =0.90$, and the right panel shows the result of the deconfinement phase (broken phase) at $\beta=1.10$.
The temporal lattice size $N_t$ is fixed to be 4 and four spatial lattice sizes $N_s$ are adopted. 
Purple, blue, green and red lines are $N_s=12$, 16, 24 and 32, respectively.

The Polyakov loop is an order parameter of the deconfinement phase transition.
In the deconfinement phase, the center symmetry is spontaneous broken and 
$\langle \Omega \rangle$ should be nonzero.
However, because of the center symmetry, the probability distribution is symmetric under $U(1)$ transformation: $\Omega \to e^{i \theta} \Omega$ for an arbitrary real number $\theta$, as shown in Fig.~\ref{fig1}.
Therefore, the expectation value of $\Omega$ is always zero even in the broken phase.
The symmetry does not break in an actual simulation with finite volume.
Therefore, to discuss spontaneous symmetry breaking, it is required to break the center symmetry adding an explicit breaking term in the action.
Then, the breaking term dependence and spatial volume dependence are investigated.
If $\langle \Omega \rangle$ is non-zero in the double limit of zero breaking term and infinite volume, we identify that spontaneous symmetry breaking has occurred.

\begin{figure}[tb]
\centering
\vspace{-3mm}
\includegraphics[width=7.0cm,clip]{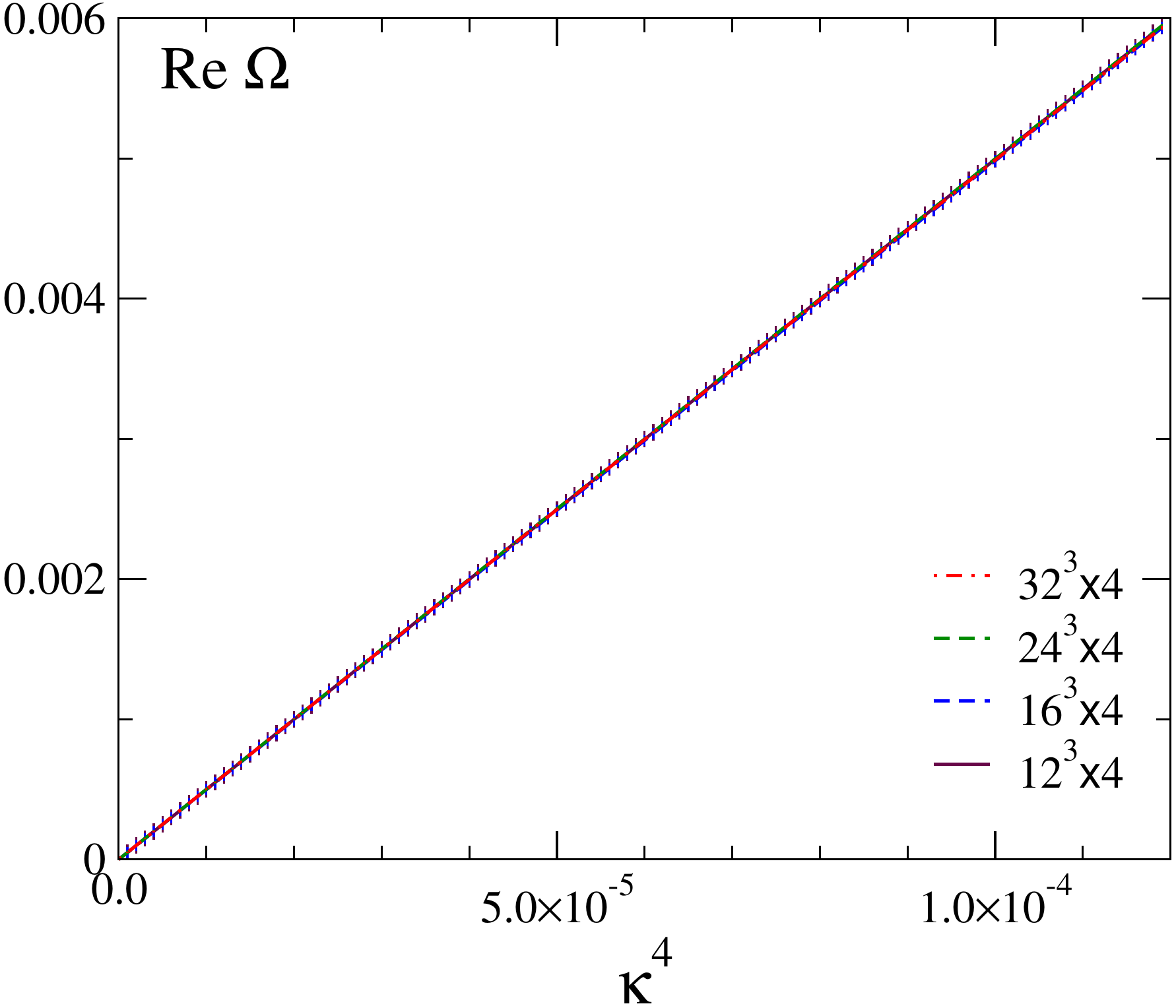}
\hspace{5mm}
\includegraphics[width=6.8cm,clip]{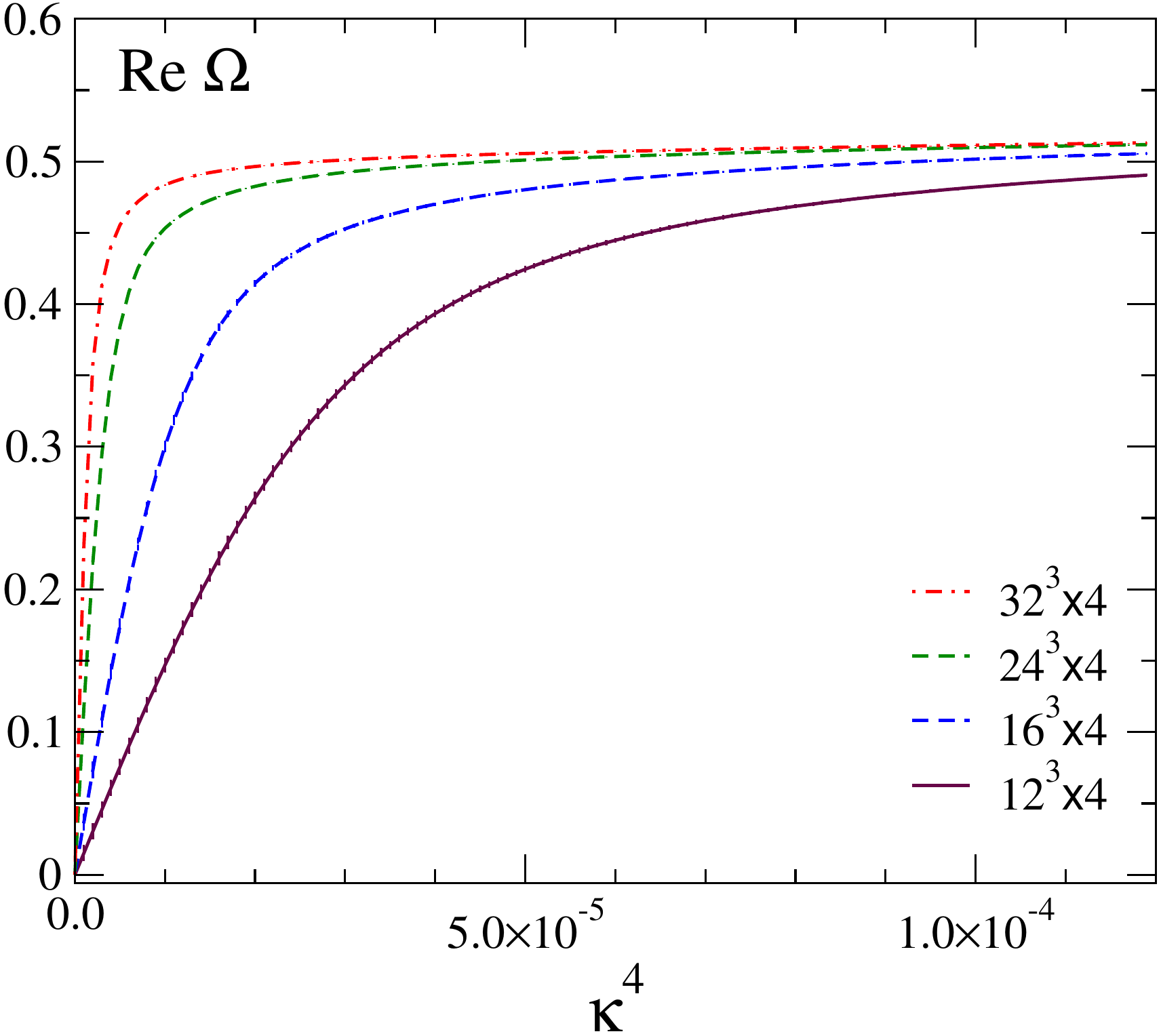}
%\vspace{-5mm}
\caption{Expectation value of the Polyakov loop as a function of the hopping parameter $\kappa$ measured at $\beta=0.90$ (left) and 1.10 (right) on $N_s^3 \times 4$ lattices with $N_s= 12$ (purple), 16 (blue), 24 (green) and 32 (red).
}
\label{fig2}
\end{figure}

To break the center symmetry, we add a heavy fermion having an infinitesimal $\kappa$.
We approximate the fermion determinant with the leading term of the hopping parameter expansion given in Eq.~(\ref{eq:proddetM}) .
The plaquette term can be absorbed into the gauge action by shifting 
$\beta \to \beta^* = \beta +16 N_{\rm f} \kappa^4$.
The expectation value is computed by the reweighting method where the reweighting factor is the Polyakov loop term,
\begin{eqnarray}
\langle {\rm Re} \Omega \rangle 
%&=& \frac{1}{Z_{GC}} \int {\cal D} U {\rm Re} \Omega \det M e^{-S_g} 
%\nonumber \\
\approx \frac{1}{Z_{GC}} \int {\cal D} U {\rm Re} \ \Omega \ e^{\epsilon V {\rm Re} \Omega} e^{-S_g} 
= \frac{\left\langle {\rm Re} \Omega \ e^{\epsilon V {\rm Re} \Omega} \right\rangle}{\left\langle e^{\epsilon V {\rm Re} \Omega} \right\rangle} ,
\end{eqnarray}
where $\epsilon =4 \times 2^{N_t} \kappa^{N_t}$, $V=N_s^3$ and 
$\langle \cdots \rangle$ means the average over quenched configurations \cite{Saito:2011fs}.
Using the data of $N_t=4$ in Fig.~\ref{fig1}, the expectation value of ${\rm Re} \Omega$ is computed.
The number of Monte-Caro updates for each $\beta$ is 200,000 for $N_s=12, 16$ and 1,000,000 for $N_s=24, 32$.
%The jackknife error is evaluated adopting an appropriate bin size.
We plot the results in Fig.~\ref{fig2} as a function of $\kappa^{N_t}$.
The left figure is the result of the confinement phase at $\beta=0.90$. 
Although the results of $N_s=12$ (purple), 16 (blue), 24 (green) and 32 (red) are plotted, no volume dependence is seen.
For this case, it is not necessary to take the volume infinity limit, thus $\langle {\rm Re} \Omega \rangle \sim \kappa^{N_t}$ and $\langle {\rm Re} \Omega \rangle$ vanishes in the $\kappa \to 0$ limit in the confinement phase (symmetric phase).
On the other hand, the right figure is the result of the deconfinement phase at $\beta=1.10$.
In the deconfinement phase (broken phase), Polyakov loop behaves as $\langle {\rm Re} \Omega \rangle \sim V \kappa^{N_t}$.
Of course, if $V$ is fixed, $\langle {\rm Re} \Omega \rangle$ becomes always zero in the limit of $\kappa = 0$.
However, in the thermodynamic limit $V \to \infty$ and $\kappa \to 0$ limit, 
this figure suggests that $\langle {\rm Re} \Omega \rangle$ becomes a non-trivial finite value in the broken phase.

Furthermore, when $\kappa$ is sufficiently small, the Polyakov loop can be evaluated by a $\kappa^{N_t}$ expansion.
Using the distribution function of the Polyakov loop in the complex plane, the expectation value can be calculated as follows, 
\begin{eqnarray}
\langle {\rm Re} \Omega^n \rangle 
&=& \frac{1}{Z_{GC}} \int {\cal D} U \ {\rm Re} \Omega^n \ e^{\epsilon V {\rm Re} \Omega} e^{-S_g} 
= \int |\Omega|^N \cos (N \phi) \ e^{\epsilon V |\Omega| \cos \phi} W(|\Omega|) \ d \phi d |\Omega|
\nonumber \\
&=& \frac{2 \pi (\epsilon V)^n}{2^n n!} \int |\Omega|^{2n} \, W(|\Omega|) \, d |\Omega| + \cdots,
\label{eq:ron}
\end{eqnarray}
where $W(|\Omega|)$ is the Polyakov loop distribution function that is independent of the complex phase and is a function of the absolute value $|\Omega|$. 
$\phi$ is the complex phase of $\Omega$.
$\langle {\rm Re} \Omega^n \rangle$ becomes zero in the limit of $\epsilon \to 0$ for finite $V$ due to the cancelation of the phase $\phi$. 
However, after integrating $\phi$, the complex phase of $\Omega$ have been already removed.
Although, in the double limit, the value of $\epsilon V$ cannot be determined, the leading term of a ratio 
$\langle {\rm Re} \Omega^4 \rangle / \langle {\rm Re} \Omega^2 \rangle^2$
does not depend on $\epsilon V$, for example.
The volume dependence of such a quantity is expected to be small.
The explicit breaking term is also important for the calculation of the canonical partition function. 
As discussed below, this method can be applied to solve the sign problem in the calculation of the derivative of the canonical partition function.

\section{Canonical partition function with a saddle point approximation}
\label{canonical}

When the probability distribution function is maximum, the chemical potential satisfies 
$-d \ln W/d \rho = -d \ln Z_C/d \rho -V \mu/T=0$, where $V=N_s^3$ and $\rho =N/V$.
Thus, the chemical potential with the maximum probability at a density $\rho$ is 
$\mu/T =-(1/V) d \ln Z_C/d \rho$. 
This derivative is given by the ratio $(d \ln Z_C/d \rho)/Z_C$. 
Both numerator and denominator are exactly zero for $N \neq 0$ due to the $U(1)$ center symmetry. 
Thus, $d \ln Z_C/d \rho$ cannot be computed without breaking the center symmetry in $U(1)$ lattice gauge theory. 
To compute $Z_C$, we apply the method we introduced in the calculation of $\langle {\rm Re} \Omega^n \rangle$ in Eq.~(\ref{eq:ron}). 

For simplicity, we calculate the canonical partition function by a saddle point approximation proposed in Ref.~\cite{Ejiri:2008xt}.
The derivative of $Z_C$ is obtained by the following equation,
\begin{eqnarray}
%- \frac{\Delta \ln Z_{\rm C} (T,N) }{\Delta N} & = & 
- \frac{1}{V} \frac{\partial \ln Z_{\rm C} (T, \rho V)}{\partial \rho} 
\approx \frac{
\left\langle z_0 \ \exp \left[ V \left( D(z_0)
- \rho z_0 \right) \right] 
e^{-i \alpha /2} \sqrt{ \frac{1}{V |D''(z_0)|}}
\right\rangle}{
\left\langle \exp \left[ V \left( D(z_0) - \rho z_0 \right) \right] 
e^{-i \alpha /2} \sqrt{ \frac{1}{V |D''(z_0)|}}
\right\rangle} .
\label{eq:chemap}
\end{eqnarray}
Here, 
$D(z)= \ln[\det M(\kappa, \mu)/\det M(\kappa_0, 0)]/V$, and 
$D''= d^2 D/dz^2 =|D''| e^{i \alpha}$. 
$z_0$ is the saddle point which satisfies $D' (z_0) =dD/d z (z_0) =\rho$. 
This bracket $\langle \cdots \rangle$ means the expected value calculated at $\kappa_0$ and $\mu=0$. 
The saddle point approximation is valid when the spatial volume is large.
The numerical computation of this equation is possible by the ordinary Monte Carlo method, 
though this calculation has the sign problem and the overlap problem.

We calculate the derivative of $Z_C$ in $U(1)$ lattice gauge theory for the case that dynamical fermions are heavy, i.e. $\kappa$ is small, and show that this calculation is possible avoiding the problem of the center symmetry and the sign problem.
To investigate the region with a small $\kappa$, we evaluate $D(z)$ by the hopping parameter expansion on each configuration,
\begin{eqnarray}
&& \hspace{-8mm} 
D(z) = 96 N_t N_{\rm f} \kappa^4 P + 2 \times 2^{N_t} N_{\rm f} \kappa^{N_t} \left[ e^z \Omega + e^{-z} \Omega^{*} \right] + \cdots , 
\label{eq:lohpe} \\
&& \hspace{-8mm} 
D' (z) = 2 \times 2^{N_t} N_{\rm f} \kappa^{N_t} \left[ e^z \Omega - e^{-z} \Omega^{*} \right] + \cdots , \hspace{5mm}
D'' (z) = 2 \times 2^{N_t} N_{\rm f} \kappa^{N_t} \left[ e^z \Omega + e^{-z} \Omega^{*} \right] + \cdots , \nonumber
\end{eqnarray}
and we perform quenched simulations with $\kappa_0 =0$.
The first term proportional to $P$ can be absorbed by the shift of $\beta$ in the quenched simulations, and the shift is very small.
We thus omit this term.
The saddle point $z_0=x_0 +iy_0$ is given by
\begin{eqnarray}
2 \times 2^{N_t} N_{\rm f} \kappa^{N_t} \left( e^{z_0} \Omega -e^{-z_0} \Omega^{*} \right) = \rho 
\label{eq:saddle}
\end{eqnarray}
on each configuration.
Because $\rho$ is a real number, 
the complex phase of $(e^{x_0+iy_0} \Omega -e^{-x_0-iy_0} \Omega^{*})$ is zero. 
The imaginary part of saddle point $y_0$ is determined as
\begin{eqnarray}
%\tan y_0 = -\frac{\mathrm{Im} \Omega}{\mathrm{Re} \Omega}, 
%\ {\rm or} \ 
y_0 = - \arctan \left( \frac{\mathrm{Im} \Omega}{\mathrm{Re} \Omega} \right) 
=- \mathrm{Arg} \Omega .
\end{eqnarray}
Then, the real part $x_0$ satisfies 
$2 \times 2^{N_t} N_{\rm f} \kappa^{N_t} \left[ e^{x_0} |\Omega| -e^{-x_0} \ |\Omega| \right] 
%=4 \times 2^{N_t} N_{\rm f} \kappa^{N_t} \sinh x_0 |\Omega|
= \rho$,
and thus 
\begin{eqnarray}
x_0 = {\rm arcsinh} \left( \frac{\rho}{4 \times 2^{N_t} N_{\rm f} \kappa^{N_t} |\Omega|} \right).
%x_0 = \ln \left( \frac{\rho} {3\times 2^{N_t+2}N_f\kappa^{N_t}} \frac{1}{|\Omega|} 
%+ \sqrt{ \left(\frac{\rho} {3\times 2^{N_t+2}N_f\kappa^{N_t}} \frac{1}{|\Omega|} \right)^2+1} \right) .
\label{eq:x0}
\end{eqnarray}
Using the saddle point approximation, the derivative of $Z_C$ is obtained by
\begin{eqnarray}
-\frac{1}{V} \frac{\partial \mathrm{ln} Z_C(T,V \rho)} {\partial \rho} 
\approx \frac{\left\langle z_0 \exp (F+i\theta) \right\rangle} {
\left\langle \exp (F+i\theta) \right\rangle}, 
\label{eq:lnzch}
\end{eqnarray}
where $F$ and $\theta$ are real numbers and are given as 
\begin{eqnarray}
F = V(D(z_0)- \rho x_0) -\frac{1}{2} \ln [V D^{\prime\prime}(z_0)] ,
%\label{eq:f} \\
\hspace{5mm}
\theta = -V \rho \ \mathrm{Arg} \Omega 
= -V \rho \ \arctan \left( \frac{\Omega_I}{\Omega_R} \right) ,
\label{eq:ftheta}
\end{eqnarray}
with
\begin{eqnarray}
D(z_0) = 96 N_t N_{\rm f} \kappa^4 P + 4 \times 2^{N_t} N_{\rm f} \kappa^{N_t} |\Omega| \cosh x_0,
\hspace{5mm}
D''(z_0) = 4 \times 2^{N_t} N_{\rm f} \kappa^{N_t} |\Omega| \cosh x_0.
\end{eqnarray}
We note that, omitting the plaquette term, $x_0$ and $F$ are functions of $|\Omega|$ and are independent of $\mathrm{Arg} \Omega$, whereas $\theta$ is a function of $\mathrm{Arg} \Omega$.
In simulations of $U(1)$ gauge theory without dynamical fermions, the probability distribution of $\Omega$ is $U(1)$ symmetric due to the center symmetry.
Thus, the numerator and denominator of Eq.~(\ref{eq:lnzch}) are exactly zero because of the phase factor $e^{i \theta}$, and Equation~(\ref{eq:lnzch}) cannot be calculated.
This is, so to say, the ultimate sign problem.

\begin{figure}[tb]
\centering
\vspace{-3mm}
\includegraphics[width=6.8cm,clip]{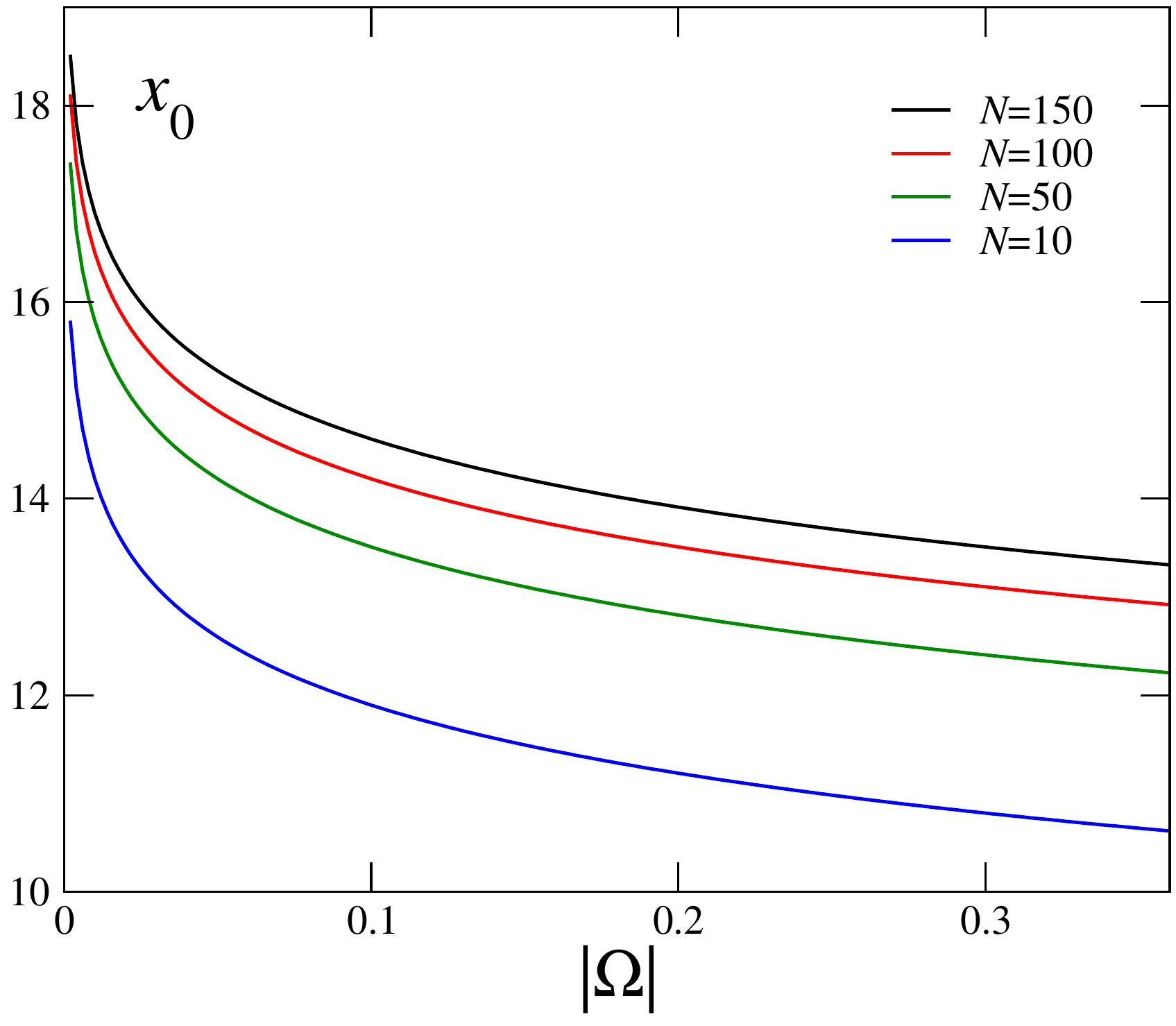}
\hspace{5mm}
\includegraphics[width=7.0cm,clip]{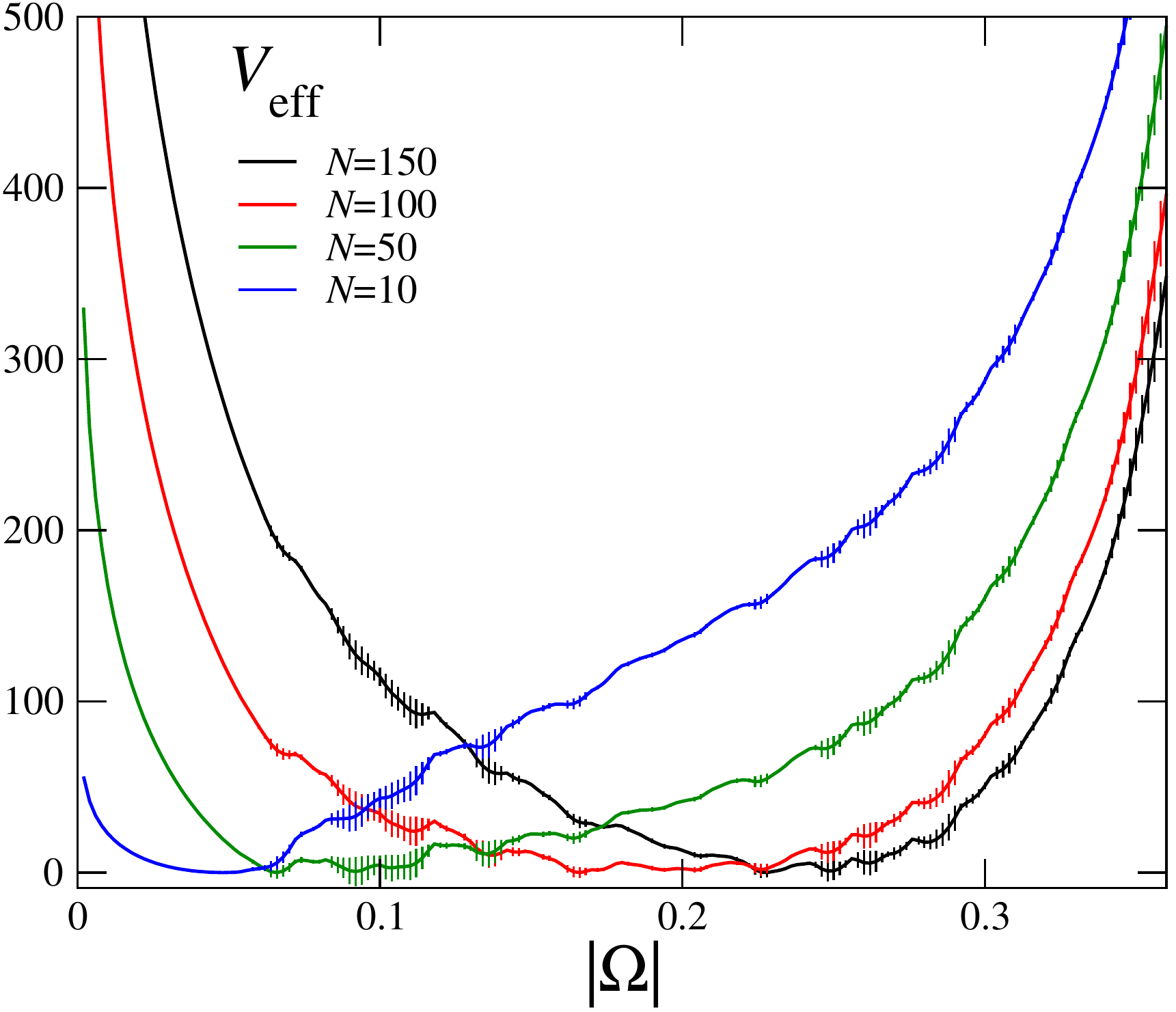}
%\vspace{-5mm}
\caption{Left: Real part of the saddle point $x_0$ as a function of the absolute value of Polyakov loop for $N=10$, 50, 100 and 150. 
Right: Effective potential of the Polyakov loop defined in Eq.~(\ref{eq:veff}) at $\beta=1.00$.
}
\label{fig3}
\end{figure}

This problem is solved by adding an additional fermion with a large mass, as in the case of discussing the expectation value of Polyakov loop.
As we calculated Eq.~(\ref{eq:ron}), we consider the $U(1)$ symmetric distribution function of the Polyakov loop and integrate $\cos(N \phi)$ with respect to the complex phase $\phi={\rm Arg} \Omega$. 
Then, 
$\int \cos (N \phi) \, e^{\epsilon V |\Omega| \cos \phi} d \phi
= 2 \pi \, (\epsilon V)^N |\Omega|^{N} /(2^N N! ) + \cdots $.
Thus, for a small $\epsilon$, the derivative of $Z_C$ becomes 
\begin{eqnarray}
-\frac{1}{V} \frac{\partial \ln Z_C(T,V \rho)}{\partial \rho} 
\approx 
\frac {\int x_0 \, e^F W(|\Omega|) \frac{2 \pi}{2^N N!} (\epsilon V)^N |\Omega|^N d|\Omega|}{\int e^F W(|\Omega|) \frac{2 \pi}{2^N N!} (\epsilon V)^N |\Omega|^N \ d|\Omega| }
=\frac {\int x_0 \, e^{-V_{\rm eff}}  \ d|\Omega| } {\int  e^{-V_{\rm eff}}  \ d|\Omega| } ,
\label{eq:dzcu}
\end{eqnarray}
where the effective potential $V_{\rm eff}  (|\Omega|)$ is defined as 
\begin{eqnarray}
V_{\rm eff} (|\Omega|) = -\ln W(|\Omega|) -F -N \ln |\Omega|.
\label{eq:veff}
\end{eqnarray}
The factor $(\epsilon V)^N$ cancels in Eq.~(\ref{eq:dzcu}).
Here, a term of $\int y_0 \sin (N \phi) d \phi$ is omitted, since this term depend on the upper bound and lower bound of the integral of $\phi$ and vanishes if we impose an appropriate dumping factor in the limit of $\phi \to \pm \infty$.

We perform simulations of $U(1)$ lattice gauge theory with the standard Wilson gauge action at several inverse gauge couplings $\beta=1/g^2$
near the deconfining transition point $\beta_c$. 
The lattice size is $N_s^3 \times N_t= 24^3 \times 6$.
Using a pseudo heat bath algorithm, the configurations are generated at thirteen $\beta$ values in the range from $\beta= 1.0000$ to $1.0240$.
The data are taken until 1,000,000 heat-bath sweeps at each $\beta$.
The multipoint reweighting method are used to combine the data generated at different $\beta$.
%The statistical errors are estimated by the jack-knife method with the bin size chosen such that the errors are saturated.
The phase transition at $\beta \approx 1.0096$ is very weak first order transition, where two phases coexist.
We adopt $N_{\rm f}=2$ and $\kappa^6 =1.92 \times 10^{-10}$.
The real part of saddle point $x_0$ for each $|\Omega|$ and $N= \rho V$ is given by Eq.~(\ref{eq:x0}), which is plotted in the left panel of Fig.~\ref{fig3}.

To compute the Polyakov loop distribution function $W(|\Omega|)$, we use a Gaussian approximation for the delta function: 
$\delta (x) \approx \exp [-(x/\Delta)^2]/(\Delta \sqrt{\pi})$ with  $\Delta =0.0025$.
%We choose the width parameter $\Delta$ by considering a balance between the resolution of the distribution function and its statistical error. 
Also, $F$ of Eq.~(\ref{eq:ftheta}) is calculated from only $|\Omega|$ and $\rho$. 
The result of the effective potential $V_{\rm eff} (|\Omega|)$ in Eq.~(\ref{eq:dzcu}) is plotted in the right panel of Fig.~\ref{fig3} for $N=10, 50,100,$ and $150$ when $\beta=1.00$.
When the spatial volume is sufficiently large, Equation~(\ref{eq:dzcu}) means that $-(1/V) d \ln Z_C/d \rho$ is the value of $x_0$ at $|\Omega|$ where the effective potential $V_{\rm eff} (|\Omega|)$ is the minimum. 
The minimum point increases as $N$ increases. 
On the other hand, $x_0$ increases as $N$ increases and decreases as $|\Omega|$ increases. 

\begin{figure}[tb]
\centering
\vspace{-3mm}
\includegraphics[width=8.0cm,clip]{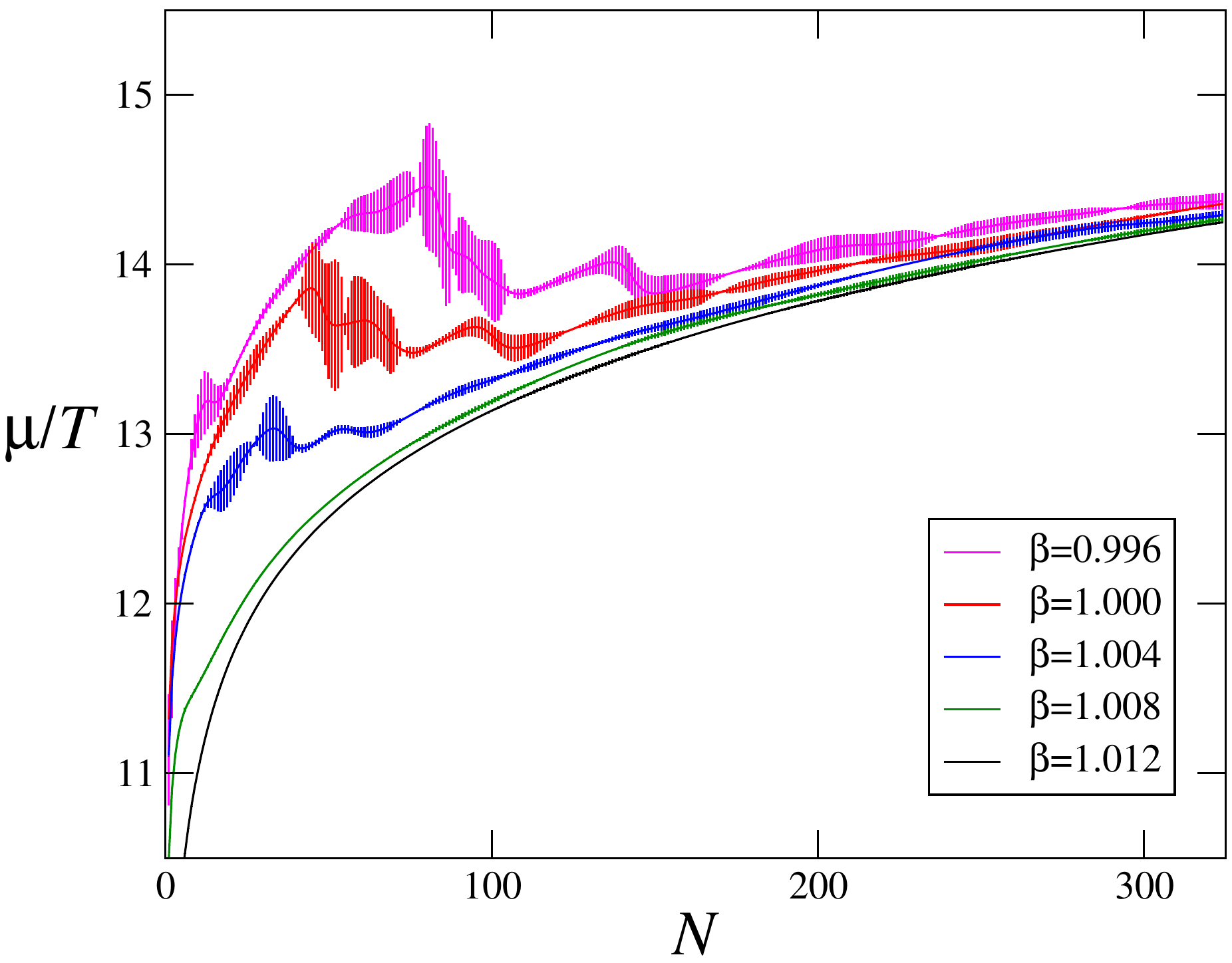}
%\vspace{-5mm}
\caption{Chemical potential $\mu /T$ for which the number of particles with the maximum generation probability is $N$ for each $\beta$ in $U(1)$ lattice gauge theory. 
}
\label{fig4}
\end{figure}

We then calculate the derivative of the canonical partition function. 
In the actual calculation, we remove $W(|\Omega|)$ as $W(|\Omega|) d |\Omega| \to {\cal D}U$ so as not to use the approximate delta function.
$-(1/V) d \ln Z_C/d \rho$
is $\mu/T$ where the density with the maximum generation probability is $\rho$, and is computable without the sign problem.
We plot the result in Fig.~\ref{fig4} for $\beta=0.996$ -- $1.012$.
When $\beta$ is in the deconfined phase at zero density, the chemical potential is monotonically increasing.
However, in the case of the confined phase at zero density, as the density increases, the chemical potential drops once and increases again.
This means that there are multiple $N$'s for a given $\mu/T$.
This is a sign when crossing the first-order phase transition.

\section{Conclusions}
\label{summary}

We studied the probability distribution function of particle density. 
The probability distribution function is obtained by constructing the canonical partition function by fixing the number of particles from the grand partition function. 
However, if the system has the center symmetry on a finite lattice, 
the canonical partition function is zero when the number of particles is not a multiple of 3 for $SU(3)$ gauge theory and when the number of particles is not zero for $U(1)$ gauge theory.
Thus, the probability distribution function is zero in these cases.
This situation is natural in the confined phase, but is unacceptable in the deconfinement phase because there should be states of various particle numbers.

This problem is essentially the same as the problem that the expectation value of the Polyakov loop is always zero when calculating with finite volume, as long as the center symmetry is not broken. 
To solve this problem, it is necessary to add an infinitesimal external field to break the symmetry and take the limit of infinite volume.
Moreover, in the case of $U(1)$ gauge theory, the sign problem can be solved using the $U(1)$ center symmetry at the same time.

We performed numerical simulations of $U(1)$ lattice gauge theory near the deconfinement phase transition point.
When the dynamical fermions are heavy, we actually demonstrated that the calculation of the probability distribution function at finite density is possible using the method proposed in this study. 
We calculated the derivative of the canonical partition function using a saddle point approximation \cite{Ejiri:2008xt}, and found that our method to avoid the sign problem works well.
The derivative is equal to $\mu /T$ for which the density of particles with the maximum generation probability is $\rho$. 
From the results of $\rho$ vs.~$\mu /T$, changes in the nature of the phase transition can be investigated.
The application of this method to QCD ($SU(3)$ gauge theory) is a future work.

%\vspace{5mm}
\subsection*{Acknowledgments} 
%\paragraph{Acknowledgments} 
The author thanks the members of the WHOT-QCD Collaboration for useful discussions.
This work was supported by JSPS KAKENHI Grant Numbers    JP21K03550, JP20H01903,
JP19H05146, and the Uchida Energy Science Promotion Foundation.
%The author also thanks the Yukawa Institute for Theoretical Physics at Kyoto University for the workshops YITP-W-19-09.

\end{document}